\documentclass[reprint,amsmath,amssymb,aps]{revtex4}
\usepackage{graphicx}
\usepackage{dcolumn}
\usepackage{bm}
\usepackage[english]{babel}
\begin{document}

\title{The Velocity Tensor and the Momentum Tensor}
\author{Tomasz Lanczewski}
\email{tomasz.lanczewski@ifj.edu.pl}
\affiliation{H. Niewodnicza\'nski Institute of Nuclear Physics, Polish Academy of Sciences\\
    Radzikowskiego 152, PL 31342 Krak\'ow, Poland}

\begin{abstract}
  \textbf{Abstract:} This paper introduces a new object called \emph{the momentum tensor}. Together with \emph{the velocity tensor} it forms a basis for establishing the tensorial picture of classical and relativistic mechanics. Some properties of the momentum tensor are derived as well as its relation with the velocity tensor. For the sake of clarity only two-dimensional case is investigated. However, general conclusions are also valid for higher dimensional spacetimes.
  \begin{description}
    \item[PACS numbers] 45.20.D-, 03.30.+p
  \end{description}
\end{abstract}

\maketitle
\section{Introduction}
  In \cite{KapLan}, an object called \emph{the velocity tensor} $V_{\nu}^{\mu}\left(\bm{v}\right)$ was described. It comes from a generalization of the equation
    \begin{equation}
        \label{1}d\bm{x}\left(t\right)-\bm{v}\left(t\right)dt=0
    \end{equation}
  into a generally covariant form
    \begin{equation}
        \label{2}V_{\nu}^{\mu}\left(\bm{v}\right)dx^{\nu}=0.
    \end{equation}
  The two-dimensional matrix of the classical velocity tensor takes the form
    \begin{equation}\label{3}
        \bm{V}(v) =V^{0}_{1}\left(
            \begin{array}{cc}
                -v & 1 \\
                -v^2 & v
            \end{array}\right),
    \end{equation}
  while in the relativistic case
    \begin{equation}\label{4}
        \bm{V}(\beta)=
        \gamma^{2}V^{0}_{1}
        \left(\begin{array}{cc}
        -\beta & 1 \\
        -\beta^{2} & \beta
        \end{array}\right),
    \end{equation}
  where $V^{0}_{1}$ is some arbitrary constant, $\beta=v/c$ and $\gamma=\left(1-\beta^{2}\right)^{-1/2}$.

  As was shown in \cite{KapLan}, the tensorial description has an obvious advantage over a standard description since it does not use the notion of the proper time
    \begin{equation}
        \label{5}d\tau =dt\sqrt{1-\frac{\bm{v}^2\left(t\right)}{c^2}}
    \end{equation}
  and therefore it allows a description of non-uniform motions and systems with an arbitrary number of material points. It also provides a cornerstone for formulating a generally covariant mechanics. However, the velocity tensor deals solely with kinematical issues. To make the tensor description complete we need to introduce another tensorial object called \emph{the momentum tensor} $\Pi^{\mu}_{\nu}(\bm{v})$.
  By means of this tensor it is possible to solve dynamical problems.
\section{Definition of the Momentum Tensor}
  In classical and relativistic mechanics the following formula holds true \cite{Gold}:
    \begin{equation}
        \frac{d\bm{p}(\bm{x},t)}{dt}=\bm{F}(\bm{x},t)\label{6}.
    \end{equation}
  The tensorial equivalent of Eq.~(\ref{6}) is presumed to be
    \begin{equation}\label{7}
        \partial_{\mu}\Pi^{\mu}_{\nu}(\bm{x},t)=\Phi_{\nu}(\bm{x},t),
    \end{equation}
  where $\Pi^{\mu}_{\nu}(\bm{x},t)$ is the momentum tensor and $\Phi_{\nu}(\bm{x},t)$ is an influence of the exterior on a body. It should be stressed here that we do not assume \emph{a priori} the relationship between $\bm{F}(\bm{x},t)$ and $\Phi_{\nu}(\bm{x},t)$. The choice of the form of the mixed tensor $\Pi^{\mu}_{\nu}(\bm{x},t)$ comes from the assumption that the momentum tensor should be some function of the velocity tensor. Since the velocity tensor is a function of a classical velocity $\bm{v}$, the momentum tensor is $\Pi^{\mu}_{\nu}(\bm{x},t):=\Pi^{\mu}_{\nu}(\bm{v})$.

\section{General Construction of the Momentum Tensor}
  In general, the momentum tensor $\Pi^{\mu}_{\nu}(\bm{v})$ is represented by a square matrix
    \[
        \bm{\Pi}(\bm{v})=
            \left(\begin{array}{cccc}
              \Pi^{0}_{0}(\bm{v}) & \Pi^{0}_{1}(\bm{v}) & \cdots & \Pi^{0}_{n}(\bm{v}) \\
              \Pi^{1}_{0}(\bm{v}) & \Pi^{1}_{1}(\bm{v}) & \cdots & \Pi^{1}_{n}(\bm{v}) \\
              \vdots           & \vdots           & \ddots & \vdots       \\
              \Pi^{n}_{0}(\bm{v}) & \Pi^{n}_{1}(\bm{v}) & \cdots & \Pi^{n}_{n}(\bm{v})
            \end{array}\right),
    \]
   where the elements $\Pi^{\mu}_{\nu}(\bm{v})$ are some functions of velocity $\bm{v}$ variable with time. In order to determine them, we make use of the transformation relation for a mixed tensor. Passing from an inertial reference frame $S$ to an inertial system $S'$ that moves with velocity $\bm{u}$ relative to $S$, the momentum tensor $\Pi^{\mu}_{\nu}(\bm{v})$ transforms in accordance with the following formula
    \begin{equation}\label{8}
        \Pi^{\mu}_{\nu}(\bm{v})\rightarrow \Pi^{\mu'}_{\nu'}(\bm{v'})=L^{\mu'}_{\mu}(\bm{u})\Pi^{\mu}_{\nu}(\bm{v})L^{\nu}_{\nu'}(\bm{u}),
    \end{equation}
  or in matrix notation
    \begin{equation}\label{9}
        \bm{\Pi}(\bm{v})\rightarrow \bm{\Pi}'(\bm{v'})=\bm{L}(\bm{u})\bm{\Pi}(\bm{v})\bm{L}(-\bm{u}).
    \end{equation}
  Assuming that $\bm{\Pi}(\bm{v})$ is form-invariant, i.e.~$\bm{\Pi}'(\bm{v'})=\bm{\Pi}(\bm{v'})$,
  we arrive at a functional equation for $\bm{\Pi}(\bm{v})$ in the form
    \begin{equation}\label{10}
        \bm{\Pi}(\bm{v'})=\bm{L}(\bm{u})\bm{\Pi}(\bm{v})\bm{L}(-\bm{u}),
    \end{equation}
  where $\bm{v'}$ is the velocity of a material point in the system $S'$ and $\bm{v}$ is its velocity in $S$. It is easy to prove \cite{KapLan} that after some simple substitutions and rearrangements in Eq.~(\ref{10}) we get the solution
    \begin{equation}\label{11}
        \bm{\Pi}(\bm{v})=\bm{L}(-\bm{v})\bm{\Pi}(\bm{0})\bm{L}(\bm{v}),
    \end{equation}
  where $\bm{\Pi}(\bm{0})$ is an arbitrary square matrix formed by constant elements.
\section{Two-dimensional Momentum Tensor}
  \subsection{Non-Relativistic Case}
    In this case we substitute in Eq.~(\ref{11}) the Galilean transformation in the form
        \[
            \bm{G}(v)=
                \left(\begin{array}{cc}
                1 & 0 \\
                -v & 1
            \end{array}\right)
        \]
    and hence we get
        \begin{equation}\label{tp5}
              \bm{\Pi}(v) =
            \left(\begin{array}{cc}
              1 & 0 \\
              v & 1
            \end{array}\right)
            \left(\begin{array}{cc}
              \Pi^{0}_{0} & \Pi^{0}_{1} \\
              \Pi^{1}_{0} & \Pi^{1}_{1}
            \end{array}\right)
            \left(\begin{array}{cc}
              1 & 0 \\
              -v & 1
            \end{array}\right)=
            \left(\begin{array}{cc}
              \Pi^{0}_{0}-v\Pi^{0}_{1} & \Pi^{0}_{1} \\
              \Pi^{1}_{0}+v\left(\Pi^{0}_{0}-\Pi^{1}_{1}\right)-v^{2}\Pi^{0}_{1} & \Pi^{1}_{1}+v\Pi^{0}_{1}
            \end{array}\right),
        \end{equation}
    where all elements $\Pi^{\mu}_{\nu}$ in Eq.~(\ref{tp5}) are constant. Since the above equation is only time-dependent, Eq.~(\ref{7}) leads to the expression
        \begin{equation}\label{tppartial}
            \partial_{0}\Pi^{0}_{\nu}(v)=\Phi_{\nu},
        \end{equation}
    where $\partial_{0}=d/dt$, and therefore we get
        \begin{equation}\label{fi0}
            \Phi_{0}=\partial_{0}\Pi^{0}_{0}(v)=\partial_{0}\left(\Pi^{0}_{0}-v\Pi^{0}_{1}\right)=-\dot{v}\Pi^{0}_{1}
        \end{equation}
    and
        \begin{equation}\label{fi1}
            \Phi_{1}=\partial_{0}\Pi^{0}_{1}(v)=\partial_{0}\Pi^{0}_{1}=0.
        \end{equation}
    Hence, in order to reconstruct the classical Newtonian equation of motion we have to assume that
        \begin{equation}\label{TPphi}
            \Pi^{0}_{1}=m\;\;\;\hbox{and}\;\;\;\Phi_{0}=-F,
        \end{equation}
    where $m$ is mass of a material point and $F$ is a classical Newtonian force in a two-dimensional spacetime. The choice of the sign in Eq.~(\ref{TPphi}) results from considerations in higher dimensional spacetimes.

    It results from Eqs.~(\ref{tp5}), (\ref{fi0}) and (\ref{fi1}) that only the element $\Pi^{0}_{1}$ takes part in dynamical processes since no other coefficient appears in Eq.~(\ref{fi0}). Therefore, the other elements may take arbitrary values and each specific choice among them will lead to the same dynamics. In particular, we may choose them in such way that the relation
        \begin{equation}\label{pfv}
            \bm{\Pi}(v)=m\bm{V}(v)
        \end{equation}
    is satisfied. Keeping in mind that $\bm{V}(v)$ is given by Eq.~(\ref{3}), we get that
        \begin{equation}\label{tp6a}
              \bm{\Pi}(v) =\Pi^{0}_{1}
            \left(\begin{array}{cc}
              -v & 1 \\
              -v^{2} & v
            \end{array}\right).
        \end{equation}
    The fact that in the considered case $\Phi_{1}=0$ leads to the general assumption that the component $\Phi_{0}$ plays a key role in the dynamics, and the components $\Phi_{k}$ are auxiliary quantities that provide the formalism covariance.

    \subsection{Relativistic Case}
    In the case of substituting into Eq.~(\ref{11}) the Lorentz transformation given by
    \[
        \bm{L}(\beta)=\gamma
        \left(\begin{array}{cc}
              1 & -\beta \\
              -\beta & 1
            \end{array}\right)
    \]
    we get that
        \begin{equation}\label{tp7}
              \bm{\Pi}(\beta) =\gamma^{2}
            \left(\begin{array}{cc}
              \Pi^{0}_{0}+\beta(\Pi^{1}_{0}-\Pi^{0}_{1})-\beta^{2}\Pi^{1}_{1} & \Pi^{0}_{1}+\beta(\Pi^{1}_{1}-\Pi^{0}_{0})-\beta^{2}\Pi^{1}_{0} \\
              \Pi^{1}_{0}+\beta(\Pi^{0}_{0}-\Pi^{1}_{1})-\beta^{2}\Pi^{0}_{1} & \Pi^{1}_{1}+\beta(\Pi^{0}_{1}-\Pi^{1}_{0})-\beta^{2}\Pi^{0}_{0}
            \end{array}\right).
        \end{equation}
    According to Eq.~(\ref{tppartial}) we obtain that
        \begin{eqnarray}\label{TPsilarel}
            \Phi_{0} & = & \partial_{0}\Pi^{0}_{0}(\beta)=\partial_{0}\gamma^{2}\left[\Pi^{0}_{0}+\beta(\Pi^{1}_{0}-\Pi^{0}_{1})-\beta^{2}\Pi^{1}_{1}\right]\nonumber \\
                     & = & \gamma^{4}\dot{\beta}\left[\left(1+\beta^{2}\right)\left(\Pi^{1}_{0}-\Pi^{0}_{1}\right)
                                                +2\beta\left(\Pi^{0}_{0}-\Pi^{1}_{1}\right)\right], \\ \nonumber \\
            \Phi_{1} & = & \partial_{0}\Pi^{0}_{1}(\beta)=\partial_{0}\gamma^{2}\left[\Pi^{0}_{1}+\beta(\Pi^{1}_{1}-\Pi^{0}_{0})-\beta^{2}\Pi^{1}_{0}\right]\nonumber \\
                     & = & \gamma^{4}\dot{\beta}\left[\left(1+\beta^{2}\right)\left(\Pi^{1}_{1}-\Pi^{0}_{0}\right)
                                                +2\beta\left(\Pi^{0}_{1}-\Pi^{1}_{0}\right)\right].
        \end{eqnarray}
    As we can observe, generally all coefficients $\Pi^{\mu}_{\nu}$ take part in the dynamics in this case since all of them are present in Eq.~(20).

    In order to illustrate the role of parameters $\Pi^{\mu}_{\nu}$ let us consider a general case of dynamics where $\Phi_{0}=const$. After the integration of Eq.~(\ref{TPsilarel}) we find that
        \begin{equation}\label{kwadrat}
            \gamma^{2}\left[\Pi^{0}_{0}+\beta(\Pi^{1}_{0}-\Pi^{0}_{1})-\beta^{2}\Pi^{1}_{1}\right]=\Phi_{0}t+C,
        \end{equation}
    where $C$ is an integration constant. Taking into consideration the initial condition for $t=0$ we obtain that
        \[
            C=\gamma_{0}^{2}\left[\Pi^{0}_{0}+\beta_{0}(\Pi^{1}_{0}-\Pi^{0}_{1})-\beta_{0}^{2}\Pi^{1}_{1}\right],
        \]
    where $\beta_{0}$ and $\gamma_{0}$ are the values for $t=0$. If we additionally assume that $\beta_{0}=0$ (i.e.~$\gamma_{0}=1$) then $C=\Pi^{0}_{0}$. Substituting this into Eq.~(\ref{kwadrat}) and making simple rearrangements we arrive at the following:
        \begin{equation}\label{rowkwad}
            \beta^{2}\left(\Phi_{0}t+\Pi^{0}_{0}-\Pi^{1}_{1}\right)+\beta\left(\Pi^{1}_{0}-\Pi^{0}_{1}\right)-\Phi_{0}t=0.
        \end{equation}
    The solutions of the above equation are of the form
    \begin{equation}\label{betarel4}
            \beta_{\pm} = \frac{\left(\Pi^{0}_{1}-\Pi^{1}_{0}\right)\pm\sqrt{\left(\Pi^{0}_{1}-\Pi^{1}_{0}\right)^{2}
                            +4\Phi_{0}t\left(\Phi_{0}t+\Pi^{0}_{0}-\Pi^{1}_{1}\right)}}{2\left(\Phi_{0}t+\Pi^{0}_{0}-\Pi^{1}_{1}\right)}.
        \end{equation}
    In the standard formalism of the Special Theory of Relativity \cite{LL}, when a constant force $F$ is applied to a body one gets the following solutions of the equations of motion for a velocity:
        \begin{equation}\label{betarelstan}
            \beta_{\pm}^{STR} = \pm\sqrt{\frac{F^{2}t^{2}}{m^{2}c^{2}+F^{2}t^{2}}}.
        \end{equation}
     If we expect that Eq.~(\ref{rowkwad}) also has two symmetric solutions, we have to assume that $\Pi^{0}_{1}=\Pi^{1}_{0}.$ Hence in this case we find that
        \begin{equation}\label{betarel5}
            \beta_{\pm} = \pm\sqrt{\frac{
                            \Phi_{0}t}{\Phi_{0}t+\Pi^{0}_{0}-\Pi^{1}_{1}}}.
        \end{equation}
     \newline
     \begin{figure}[hb]
        \begin{center}
        \includegraphics[width=0.5\textwidth]{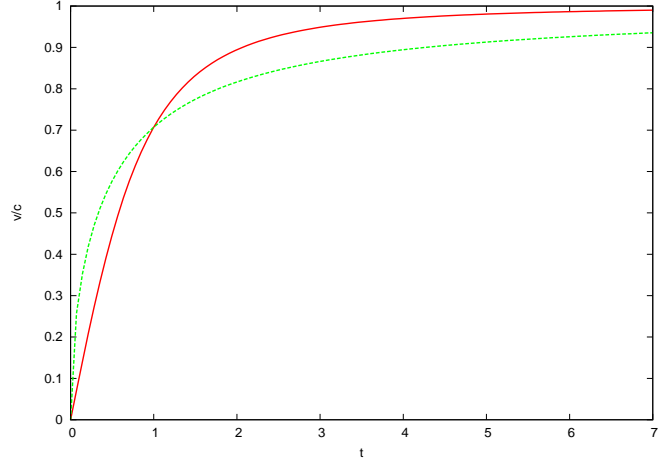}
        \caption{Comparison of $\beta(t)$ (green) and $\beta^{STR}(t)$ (red). $F=1$, $\Phi_{0}=1$, $m^{2}c^{2}=\Pi^{0}_{0}-\Pi^{1}_{1}=1$ are assumed here.}
     \end{center}
     \end{figure}\newline
     It should be stressed here that the asymptotes of Eqs.~(\ref{betarelstan}) and (\ref{betarel5}) are identical, i.e.:
        \[
            \lim_{t\rightarrow\infty}\beta_{\pm}^{STR}=\lim_{t\rightarrow\infty}\beta_{\pm}=\pm 1\;\;\;\hbox{and}\;\;\;
            \lim_{t\rightarrow 0}\beta_{\pm}^{STR}=\lim_{t\rightarrow 0}\beta_{\pm}=0.
        \]
     As we can see from Eq.~(\ref{betarel5}), the constant $\Pi^{1}_{1}$ plays the role of a ``renormalization'' constant for $\Pi^{0}_{0}$, hence it can be discarded without losing the generality of considerations. Then matrix (\ref{tp7}) takes the form
        \begin{equation}\label{TP1STW2}
            \bm{\Pi}(\beta) =\gamma^{2}
            \left(\begin{array}{cc}
              \Pi^{0}_{0} & \Pi^{0}_{1}-\beta\Pi^{0}_{0}-\beta^{2}\Pi^{0}_{1} \\
              \Pi^{0}_{1}+\beta\Pi^{0}_{0}-\beta^{2}\Pi^{0}_{1} & -\beta^{2}\Pi^{0}_{0}
            \end{array}\right).
        \end{equation}
     Matrix (\ref{TP1STW2}) can also be rewritten as
        \begin{equation}\label{TP1STW3}
            \bm{\Pi}(\beta) =\gamma^{2}\Pi^{0}_{0}
            \left(\begin{array}{cc}
              1 & -\beta \\
              \beta & -\beta^{2}
            \end{array}\right)+
            \Pi^{0}_{1}
            \left(\begin{array}{cc}
              0 & 1 \\
              1 & 0
            \end{array}\right),
        \end{equation}
     where the second matrix on the right hand side of Eq.~(\ref{TP1STW3}) is constant in time.

    Assuming that $\Pi^{0}_{1}=\Pi^{1}_{0}$ and $\Pi^{1}_{1}=0$, Eqs.~(\ref{TPsilarel}) and (21) turn into
        \begin{eqnarray}\label{TPsilarel2}
            \Phi_{0} & = & \partial_{0}\Pi^{0}_{0}(\beta)=\partial_{0}\gamma^{2}\Pi^{0}_{0}\nonumber \\
                     & = & 2\gamma^{4}\dot{\beta}\beta\Pi^{0}_{0},\nonumber \\ \\
            \Phi_{1} & = & \partial_{0}\Pi^{0}_{1}(\beta)=\partial_{0}\gamma^{2}\left(-\beta\Pi^{0}_{0}\right)\nonumber \\
                     & = & -\gamma^{4}\dot{\beta}\left(1+\beta^{2}\right)\Pi^{0}_{0}.\nonumber
        \end{eqnarray}
    In order to compare it with the standard formalism of the Special Theory of Relativity, let us recall that in the standard description the equation of motion is given by \cite{LL}
        \[
            F=\frac{dp}{dt}=\frac{mc\dot{\beta}}{\left(1-\beta^{2}\right)^{3/2}}=\gamma^{3}mc\dot{\beta},
        \]
    and therefore
        \[
            \dot{\beta}=\gamma^{-3}\frac{F}{mc}.
        \]
    Substituting this expression into Eq.~(\ref{TPsilarel2}) we get
        \begin{eqnarray}\label{TPsilarel3}
            \Phi_{0} & = & 2\gamma\frac{F}{mc}\beta\Pi^{0}_{0},\nonumber \\  \\
            \Phi_{1} & = & -\gamma\frac{F}{mc}\left(1+\beta^{2}\right)\Pi^{0}_{0}.\nonumber
        \end{eqnarray}
    This indicates that the assumption that $\dot{\beta}$ in this formalism and the standard description is the same leads to the conclusion that for a force $F$ constant in time the component $\Phi_{0}$ is not constant in time, and \emph{vice versa}. However, the uniform motion ($\dot{\beta}=0$) in both formalisms occurs simultaneously.

    The non-trivial part of the matrix (\ref{TP1STW3}) can also be expressed by means of well-known relativistic quantities such as energy and momentum:
        \[
            E=\gamma mc^{2},\;\;\; p=\gamma mc\beta.
        \]
    Therefore we get
    \begin{equation}\label{TP3STW}
            \bm{\Pi}(\beta)=\frac{\Pi^{0}_{0}}{m^{2}c^{4}}
            \left(\begin{array}{cc}
              E^{2} &-Epc  \\
              Epc & -p^{2}c^{2}
            \end{array}\right)+
            \Pi^{0}_{1}
            \left(\begin{array}{cc}
              0 & 1 \\
              1 & 0
            \end{array}\right).
        \end{equation}

    It should be highlighted here that --- as was mentioned before --- it is possible to choose a different special form of the relativistic velocity tensor matrix and --- consequently --- a different description of dynamics. For instance, by analogy with the non-relativistic solution, we can assume that the relation between the velocity tensor described by Eq.~(\ref{4}) and the momentum tensor is given by Eq.~(\ref{pfv}).
    Hence in order to reproduce Eq.~(\ref{pfv}) the general form of the momentum tensor matrix (\ref{tp7}) has to be reduced to the matrix
        \begin{equation}
           \label{P2rel}
           \bm{\Pi}(\beta)=
            \gamma^{2}\Pi^{0}_{1}
            \left(\begin{array}{cc}
            -\beta & 1 \\
            -\beta^{2} & \beta
            \end{array}\right),
        \end{equation}
    where --- as we have shown for the non-relativistic case --- the constant $\Pi^{0}_{1}$ can be identified with mass $m$ of a material point. It is easy to observe that form (\ref{P2rel}) is obtained from Eq.~(\ref{tp7}), where all coefficients with the exception of $\Pi^{0}_{1}$ vanish. Therefore, Eqs.~(\ref{TPsilarel}) and (21) can be written down as:
        \begin{eqnarray}\label{TPsilarelX}
            \Phi_{0} & = & -\gamma^{4}\left(1+\beta^{2}\right)\dot{\beta}\Pi^{0}_{1}, \nonumber \\
            \Phi_{1} & = & 2\gamma^{4}\beta\dot{\beta}\Pi^{0}_{1}.\nonumber
        \end{eqnarray}

\section{Conclusions}
  The aim of this paper was to introduce a new dynamical object called the momentum tensor as an analogue to the kinematical velocity tensor, and therefore to complete the tensorial description of classical and relativistic mechanics. Calculations show that the choice of constants in the momentum tensor matrix results in different models of dynamics in the relativistic case. Another important fact is that the naturally assumed relation between the tensors: $\bm{\Pi}(v)=m\bm{V}(v)$ is just one among many. Further investigations will focus on verifying the other models.
\acknowledgments
I would like to thank Prof.~Edward Kapu\'scik for his scientific advice, and also for useful comments and ideas on this subject.

\end{document}